# Control of human immune response function by T-cell population fluctuation and relaxation dynamics

**Susmita Roy and Biman Bagchi***

**SSCU, Indian Institute of Science, Bangalore**

**Email:profbiman@gmail.com**

## *Abstract*

**Clinical studies have indicated that in malignant surveillances fluctuations in the population of certain effector T-cell repertoire become suppressed. Motivated by such observations and in an attempt to quantify adaptive human response to pathogens, we define an immune response function (IMRF) in terms of mean square fluctuations of T-cell concentrations. We employ a recently developed kinetic model of T-cell regulation that contains the essential immunosuppressive effects of vitamin-D. We employ Gillespie algorithm to make the first study of fluctuations along the stochastic trajectories. We show that our fluctuation-based IMRF parameter can be more robust than using only the average value of T-cell concentration or ratio of different T-cell subsets. This function can differentiate responses of different individuals after pathogenic incursion both under healthy and disease conditions. We find that relative fluctuations in T-cells (and hence IMRF) are different in strongly regulated (malignant prone) and weakly regulated (autoimmune prone) regions. *The cross-over from one steady state (weakly regulated) to the other (strongly regulated) is accompanied by a divergence-like growth in the fluctuation of both the effector and regulatory T-cell concentration over a wide range of pathogenic stimulation, displaying a dynamical phase transition like behavior.* The growth in fluctuation in this desired (or, healthy) immune response regime (with high IMRF) is found to arise from an intermittent fluctuation between regulatory and effector T-cells that results in a bimodal distribution of population of each, indicating bistability. The signature of intermittent behavior is further confirmed by calculating the power spectrum of the corresponding fluctuation of time correlation function. The calculated time correlation functions of fluctuations show that while the slow fluctuation causes the bistabilty in healthy state, the same fluctuation relaxes, in contrast, at relatively faster rate in disease states. Dimmed fluctuations (equivalent to lower IMRF) can indeed be a characteristic feature of the strong or weakly regulated states. Thus, in diseases diagnosis process, such steady state response parameters can provide immense information which might become helpful to define an immune status.**

# I. Introduction

We scarcely realize that we are alive because inside our body there is a constant, highly vigilant microscopic mechanism designed to defend us every moment from millions of bacteria, microbes, viruses, fungi, parasites and pollutions. This amazing protection mechanism is called the immune system. It is indeed extremely efficient machinery that consists of a highly complex network involving billions of immune components and their interactions and chemical reactions. However efficient it is, depending on the type of perturbation, it often falls short that leads to a disease condition [1]. As a result we require external treatment to cure the disease. Hence it is rather important to diagnose what is the disease condition and what could be the suitable marker/s for in the diagnostic treatment.

Despite its importance, our understanding of human immune response has remained rather rudimentary. While there do exist clinical observations and systematic experimental studies that can help us identifying some of the essential immune ingredients and their functions [2-13], we often fail to quantify immune responses because we lack specification of sensitive order parameters that can characterize state of immunity and foretell the onset of disease. Instead, the standard practice is to look for the average values of a number of immune components and opt for their values in the standard range as signatures of healthy system. Often serious malfunction of the immune system develops over a period of time when average values remain within the range prescribed and may thus fail to give any indication of the impending crisis. Even more concerning is the fact that we take note only when the parameters have already entered the disease condition.

Additionally, one should remember that our immune response is not only highly case (that is, disease) specific but is also personalized. It not only depends on external factors like pathogenic strength, pollution, etc. but also on internal individual immune tolerance [14-16]. Therefore, we need to quantify immune response in terms of a suitable order parameter so that clinical diagnosis can more sensitively probe and foretell a disease condition. Before characterizing an appropriate marker we need to know some essential information in the regulation of immune response.

To overcome the onslaught of innumerable types of ever-evolving pathogens, some immune cells having unlimited repertoire of receptors ensure that the invading pathogens are recognized [17-20]. After the antigen-specific recognition by receptors on lymphocytes, the clonal expansion of pathogen-specific lymphocytes (in brief, specific effector T-cells), directly or indirectly clear the pathogen load. Since the receptors are generated in a random process, these effector immune cells often fail to distinguish self from non-self. As a result, it leads to a serious threat to health which is the development of autoimmunity [20-23]. A crucial prerequisite is the elimination of such self-reactive receptors by the mechanism of self-tolerance [14-16].

Although both genetic and environmental factors affect disease prevalence, autoimmunity is primarily driven by the enhanced population of Th1 helper cells that attack various self-tissues in the body [24, 25]. A number of other effector T cell subsets, like Th17 and Th9 cells, and regulatory T cells, like Tregs and Tr1 cells, also strongly participate in autoimmunity at cellular and molecular levels. Although IL-17 producing Th17 cells are believed to serve as one of efficient players in autoimmunity, the intricate balance between $T_{Reg}$s and Th17 cells, provide a basis for understanding the immunological mechanisms that induce and regulate autoimmunity [20]. While effector T cells control inflammation, regulatory T cells play a vital role to control it. Physicians usually treat autoimmune diseases with an immunosuppressive drug that decreases the activity of the effector T-cells so that it does not harm body's own tissues or transplanted organs or tissues. Like corticosteroid medications, vitamin-D is also being exploited as an excellent immune modulator for the prevention of autoimmune disorders [26, 27].

Overactive immune-suppression often may edit the immune response in such a way that malignant cells may find it relatively easier to evade the immune system [28-30]. This immune-evasion has been identified as a hall-mark on the onset of at least some types of cancer. Several reports documented that the progression of cancer is involved with an enhanced activity of regulatory T-cells and decreasing effects of effector T-cells [31-38]. This enhanced activity of regulatory T-cells in presence of excess level of vitamin-D causes significant annihilation of effector T-cell population. Thus the opposing role of regulatory and effector T-cells in immunological activity often determine the strength of immune-regulation and the fate of a disease [20]. The regulation is largely determined by the activation of APCs followed by the production of effector T-cells [39-43].

Based on such T-cell response, we have categorized the regulation into three groups: (i) strong regulation, where the population of effector T-cell is less in comparison to the population of regulatory T-cell; (ii) weak regulation, where effector T-cells are larger in population than the regulatory T-cells; and (iii) moderate regulation where both effector and regulatory T-cells maintain a balanced population [44, 45]. We show below that the balance is maintained by a bistability.

In our earlier study we developed a coarse grained but general kinetic model in an attempt to capture the role of vitamin-D in immunomodulatory responses [44]. We found that although vitamin-D plays a negligible role in the initial immune response, it exerts a profound influence in the long term, especially in helping the system to achieve a new, stable steady state. The study explores the role of vitamin-D in preserving an observed bistability in the phase diagram (spanned by system parameters) of immune regulation, thus allowing the response to tolerate a wide range of pathogenic stimulation which could help in resisting autoimmune diseases [44].

To investigate the above mentioned regulation limits we have performed time evolution analysis of each participating element after the pathogenic attack to study their long time steady state behavior. Most of the clinical and experimental studies use some standard values of effector and regulatory T-cell, more often the ratio of effector T-cells to $T_{Reg}$s as the predictor of patients' immune profile [46, 47]. However, such standardized ratios or numbers are frequently found to be challenged by other related studies or experimental results. Often we ignore that these standard numbers can be personalized and it may vary with patients even carrying the same disease. Under such circumferences both experimental and theoretical investigations should find suitable parameters that can quantify person specific immune response. With such objective in mind, we develop and solve a coarse grained kinetic network model to probe and quantify the sensitive immune response. In our model, person specific information enters through a set of rate constants (like production rates of effector and regulatory T-cells). The state of a person's immune system is also quantified by the concentration of inactive vitamin-D and other naturally occurring immunomodulatory chemicals/hormones and also concentration of native T-cells, among other things. Although varying within a range, these values differ from person-to-person.

It is difficult to get a detailed quantitative map of many of these parameters and also relationship of these values to the immune response function of an individual. A theoretical model can go a long way in understanding relationship between concentrations of various species and immune response.

In many cases, a pre-disease condition may last for a long time. The average values of concentrations of lymphocytes (T-cells) may not offer any clue that a person is in the pre-disease condition. However, there may be certain other order parameters that may be able to capture the disease condition. We propose here and demonstrate *that fluctuations of the T-cells can capture certain aspects of the pre-disease condition and can be a sensitive measure to follow.* A brief concept of response functions that is associated with the fluctuation phenomena in this relation we discuss below.

## II.Theoretical description of immune response

### A. Definition of immune response in terms of fluctuations

A response function provides a quantitative measure of the response of a system to external perturbation [48, 49]. In equilibrium statistical mechanics, such response functions are defined in terms of mean square fluctuations in the conjugate system properties [50-52]. For example, specific heat is expressed in terms of mean square energy fluctuation, isothermal compressibility in terms of number density (or, volume) fluctuations, dielectric constant in terms of polarization (or, total dipole moment) fluctuations. The reason for these relations between response functions and fluctuations is that the latter provides the ability of the system to absorb external perturbation. In the case of time (or frequency) dependent perturbation, the response also becomes time or frequency dependent. There is an intimate relation between the time dependent response function and a dynamical property of the system at equilibrium. This relation is expressed by linear response theory (LRT), first clearly formulated by Kubo in 1957 [50].

Kubo's landmark analysis of response function starts by splitting a Hamiltonian $(H)$ into an unperturbed $(H_0)$ and a time-dependent perturbation $H'(t)$

$$H = H_0 + H'(t)$$

Let us consider an isolated system and its Hamiltonian is denoted by $H$. The dynamical motion of the system determined by $H$ is called "*natural motion*" of the system [53]. We suppose that an external force $F(t)$ is applied to the system, the effect of which is represented by the perturbation energy,

$$H'(t) = -AF(t)$$

where $A$ is a dynamical variable that is conjugate to field *F*. For example, as we just mentioned that if $F(t)$ is an electric field, then $A$ can be the total dipole moment of the system. Similarly, if $F(t)$ is pressure, then $A$ should be the volume [54].

*We generalize the above concepts to define an immune response for a living system. In our case the external perturbation is provided by the pathogen and the conjugate property is the T-cell concentration that couples to pathogen invasion.* Therefore, we define the first immune response function (IMRF) in terms of effector T-cell concentration, defined as

$$\text{Immune response,}\ IMRF(i,j) = \frac{\left\langle (\delta[T(i,j)])^2 \right\rangle}{\left\langle [T(i,j)] \right\rangle}$$

where $\delta[T(i,j,t)] = [T(i,j,t) - \langle T(i,j)\rangle]$. Here we assume T cell concentration is a function of i and j where i represents an individual and j represents a disease.

We note that our system is not an equilibrium system. However, the system does attain a steady state, even in the presence of pathogenic load.

## B. Source of fluctuations

Let us consider a small volume element in our body, either in a blood sample or in a tissue. We consider that in that volume element there is initially only one pathogen and a certain number of native T-cells that on appropriate signal reception can convert into effector or regulatory T-cells. We also assume a fixed number of antigen presenting cells (APC) that in the presence of pathogen activates native T-cell.

As stated, there is only one pathogen at time t=0 in our volume element. Subsequent time evolution of pathogen concentration depends on various factors. (i) The rate of production of pathogen which varies from pathogen to pathogen. (ii) Efficiency of APC to detect and activate native T-cells into effector T-cells. (iii) Efficiency of the effector T-cells in destroying pathogens, (iv) generation of regulatory T-cells that involves again APC and also vitamin D, for example. These efficiencies are reflected in certain rate parameters that are different from person-to-person.

It is not emphasized sufficiently that while the rate parameters are characterized by an average value but also by a distribution. Thus, even when we start with fixed initial concentration, time evolution gives rise to trajectories that can be different in different trials. That is because distribution in rates brings in a distribution of the concentration of active participants.

Even when rate constant is kept fixed, there is fluctuation due to stochasticity of the type envisaged in Gillespie algorithm. This is the stochasticity inherent in a bimolecular reaction. In a coupled system of non-linear equations, such fluctuations can get enhanced near a dynamical crossover. It is noteworthy that while internal spontaneous statistical fluctuations depends on the system's state, the effect of external fluctuations is much more profound and modifies completely the domain in which a phase transition occurs. More about the implications of such microscopic fluctuation are discussed in the result section.

## C. Time dependence of immune response

If one considers a time dependent change in the concentration of T-cell then one can write a time dependent response as,

$$IMRF(i,j,t) = \int_0^t dt' \int_0^t dt'' \left\langle [T(i,j,t')][T(i,j,t'')] \right\rangle$$

The quantity $\left\langle [T(i,j,t')][T(i,j,t'')] \right\rangle$ is an example of time correlation function [53, 54]. This quantity measures the degree of correlation in steady state value of T-cell at two different times, $t'$ and $t''$.

All these response functions we desire to calculate from a stable kinetic network model of T-cell regulation that we developed in our early work [44].

## D. Quantification of immune response function (IMRF) : Connecting series of events

The important constituents of the model are the following: (i) pathogen (it is important to note that, in our analyses we have considered pathogen, as a numerical quantity "P" that is capable of eliciting T-cell mediated immune response), (ii) naive T-cell, (iii) myeloid dendritic cell in the form of professional antigen presenting cells (APC), both in their resting (immature) and activated (mature) forms (iv) effector and regulatory T-cells, (v) inactive vitamin-D ($25(OH)D_3$) and active vitamin-D ($1,25(OH)_2D_3$). However the participants, such as vitamin-D receptor (VDR) and the enzyme $25(OH)D_3$-1α-hydroxylase (CYP27B1) that simultaneously convert inactive vitamin-D to active vitamin-D ($1,25(OH)_2$ D)-VDR [D*-VDR] protein complex, are considered as implicit factors for activation of the required transcriptional motif [44].

It is important to emphasize here that we have essentially combined three important experimental observations those include the essential features of the adaptive responses reported by (i) Powri and Maloy, [42] (ii) Jorge Correale et al. [37] and (iii) Lorenzo Piemonti et al. [43]. In **Figure 1** we have presented the complex interaction network model that comprises various components and their inter-relation and regulation involved in the immune system.

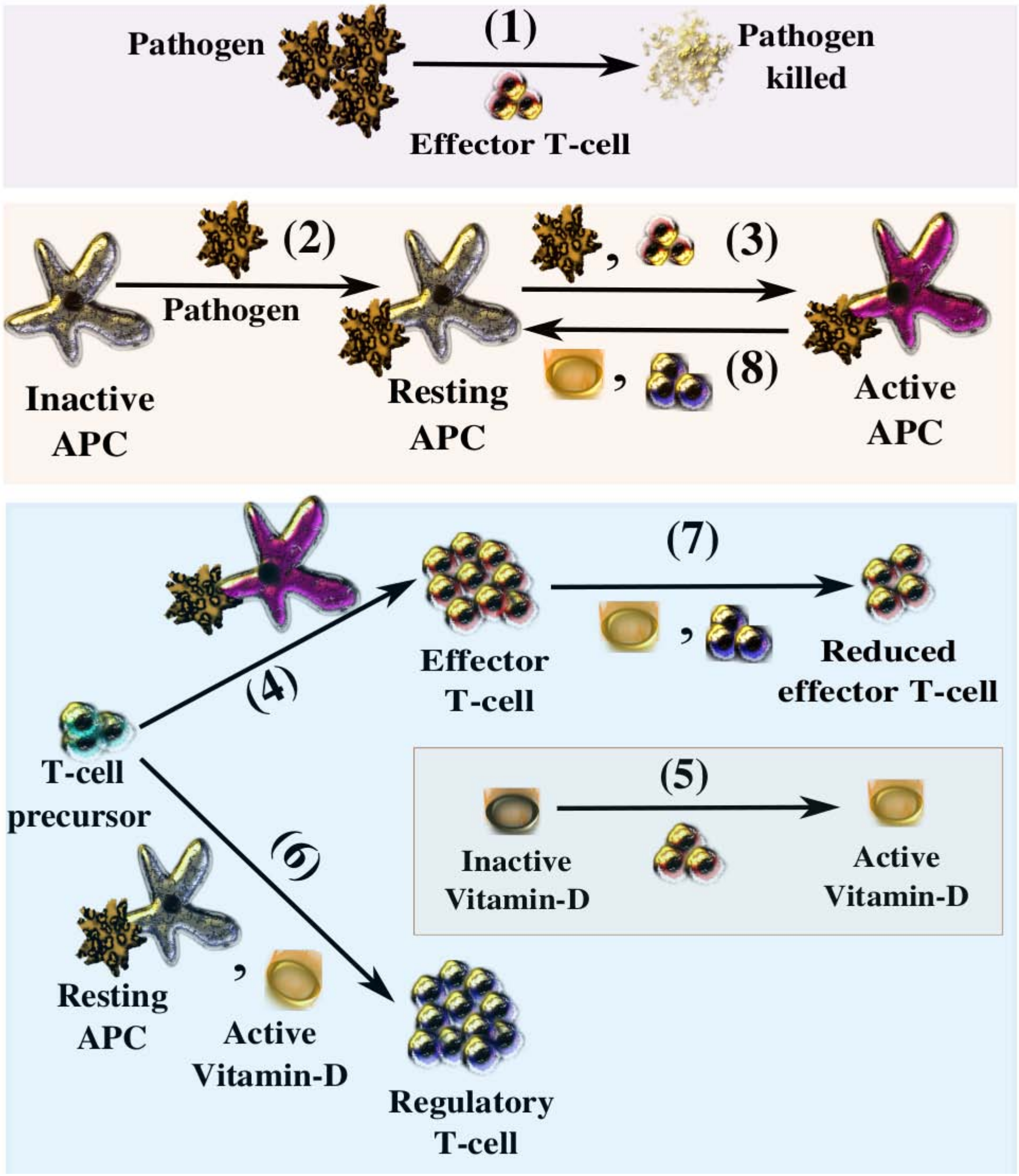

**Figure 1: A schematic representation of adaptive immune responses in terms of cellular interactions including vitamin-D, based on some experimental results and clinical observations. In the scheme, the primary events are the following: (1) The main step is the annihilation of pathogen by effector T-cells. (2) In presence of pathogen, inactive APC becomes stimulated after pathogen recognition and form resting APC. (3) Resting APC is activated either by pathogen or by the presence of any effector T-cell [41, 42]. (4) Activation of effector T-cells are initiated by these active APC. (5) Effector T-cells initiates the formation of active vitamin-D from its inactive form [36]. (6) Resting APC and active vitamin-D both can stimulate the production of regulatory T-cell from its precursor [37]. (7) Enhancement in the rate of production of effector T-cells is controlled by both**

**regulatory T-cells ($T_{Reg}$) and active vitamin-D [36]. (8) In addition, vitamin-D and regulatory T-cell up-regulate the formation of more resting APC from active APC [43].**

The present approach of network kinetic model building bears strong resemblance to similar methods adopted earlier in the study of kinetic proof reading [55, 56] and also in enzyme kinetics [57-59]. In all these studies, precise quantitative prediction is hindered by insufficient knowledge about the system parameters. Especially, the values of rate constants are often not available. This lacuna is indeed a source of serious problem not only in study of kinetic proof reading and enzyme kinetics but also, as we find here, in any theoretical investigation of immunology. Finally, master equations involved in all such problems are solved by employing the method of mean first passage time [56, 60, 61], Gillespie algorithm or straight forward numerical integration. We adopt stochastic simulation approach by using Gillespie algorithm [62, 63]. The coarse-graining of interaction network, development of the reaction scheme and the master equations are discussed in the following section.

The values of parameters involved (rate constants and concentrations) may span a wide range, and can vary from disease to disease, and from person to person. Thus, a study of the response to the variation of the important parameters has been carried out. Such a study is clearly necessary in the present context.

## III. Framework of kinetic model

### A. Coarse-grained reaction network model development

In order to describe the complex interplay among different types of immune cells, pathogens and the modulatory role of vitamin-D, first we need to develop a simple coarse-grained approach that can both be solved and understood. The complexity arises because of the large number of biochemical machineries in the human body that are strongly coupled with each other [64, 65]. Understanding the relationship between these different machineries involving different types of cell may ultimately require detailing at the molecular level. A simpler, albeit cruder version is proposed here that accounts for some of the complexities present at the molecular level by coarse-graining them at the cellular level (**Figure 1**). With this goal in mind, we perform model analyses of T-cell activation, deactivation and regulation, and establish connection with a few experimental results discussed below.

(i) Myeloid dendritic cells (also we call them as antigen presenting cell (APC)) present in different organs, are the key players involved in triggering the onset of an adaptive immune response. Upon maturation and pathogen presentation, these dendritic cells serve to activate naive T-cells into effector T-cells. In contrast, immature dendritic cells upon pathogen contact convert naive T-cells into regulatory T-cells in the absence of maturation signal [41, 42].

(ii) Effector T-cells release cytokines which upregulate the activity of 1α-hydroxylase enzyme (CYP27B1), which in turn, induces the conversion of active vitamin-D from its inactive form. The extensive experimental study by Correale et al.

reported that CD4+ T-cells are capable of metabolizing 25(OH)-vitamin D to 1,25 $(OH)_2$-vitamin D, which again inhibits the function of T-cells [37].

(iii) Active form of vitamin-D, 1, 25$(OH)_2D_3$ [D*] modulates the immune response through the inhibition of DC differentiation and maturation into potent APC [43].

(iv) Increased production of [D*], directly inhibits effector T-cell [$T_{Eff}$] production and upregulates CD4+/CD25+/FoxP3+ regulatory T-cell [$T_{Reg}$] response. These $T_{Reg}$ cells also efficiently inhibit $T_{Eff}$ cells proliferation [34-37].

Coarse-graining of the interaction network is accomplished by making several simplifying assumptions. They are as follows:

(a) Th1, Th2 and Th17 cells are grouped together as effector T-cells. The detailed description of these T-cells is depicted in our previous work [44]. That is, we assume that they act together, in a parallel fashion.

(b) It is well established that the primary molecular action of 1,25$(OH)_2D_3$ is to initiate gene transcription by binding to VDR which is a member of the steroid hormone receptor superfamily of ligand-activated transcription factors. VDR therefore is an important factor in 1,25$(OH)_2$ $D_3$ mediated functions. More detailed information about VDR can be found in ref 66 [66]. On the contrary, there are reports that 1,25$(OH)_2D_3$ also has rapid actions that are not essentially mediated through transcriptional events involving VDR. They are in fact membrane initiated actions [67]. In the present model we have not included the effect of VDR. We have only considered the production of active vitamin-D from its inactive form upon T-cell activation.

(c) In tests, the inactive form of vitamin-D, 25(OH)$D_3$, is generally used as an indication of vitamin-D status. However, in dendritic cells (DC) use of this precursor depends on its uptake by cells and subsequent conversion by the enzyme CYP27B1 into active [D*] [66]. [D*] has a tight control over the homeostatic production rate that auto-regulates its production, by directly upregulating the activity of the P450 cytochrome CYP24A1. In our model we have considered the steady state rate of inactive vitamin-D that found from experimental and clinical measurements while keeping the concentration of these enzymes as the implicit factors.

In the present analysis we consider the following set of connected biological reactions. Most of them are catalytic reaction in terms of up-regulation or down-regulation.

(1) The primary step is the annihilation of pathogen by effector T-cell.

$$\text{Pathogen }(\text{P})+\text{Effector T-cell }(\text{T}_{\text{Eff}})\rightarrow \text{P}_{\text{killed}}+\text{T}_{\text{Eff}} \qquad \text{(i)}$$

(2) Production of effector T-cell requires the presence of active antigen presenting cell (APC). Active APC, on the other hand is produced by the following sequence of reactions. $1^{st}$ resting APC forms through the interaction between inactive APC and pathogen.

$$\text{Inactive APC }(\text{A}_{\text{in}})+\text{P}\rightarrow \text{Resting APC }(\text{A}_{\text{Res}})+\text{P} \qquad \text{(ii)}$$

(3) Further pathogenic contact and/or effector T-cell contact promotes the resting APC to turn out to be active APC.

$$\left\{\begin{matrix} \text{A}_{\text{Res}}+\text{P}\rightarrow \text{A}_{\text{Act}}+\text{P} \\ \text{A}_{\text{Res}}+\text{T}_{\text{Eff}}\rightarrow \text{A}_{\text{Act}}+\text{T}_{\text{Eff}} \end{matrix}\right\} \qquad \text{(iii)}$$

(4) Then effector T-cell is produced by the interaction between precursor/naive T-cell with active APC.

$$\text{Naive T-cell } \left(T_{Na}\right) + \text{Active APC } \left(A_{Act}\right) \rightarrow T_{Eff} + A_{Act} \qquad \text{(iv)}$$

(5) Simultaneously inactive vitamin-D is transformed into active vitamin-D upon effector T-cell contact.

$$\text{Inactive Vitamin-D } \left(D_{in}\right) + T_{Eff} \rightarrow \text{ActiveVitamin-D } \left(D^*\right) + T_{Eff} \qquad \text{(v)}$$

(6) Resting T-cells and vitamin-D, both can initiate the formation of regulatory T-cell from naive T-cell.

$$\left\{\begin{array}{l} \text{Naive T-cell } \left(T_{Na}\right) + \text{Resting APC } \left(A_{Res}\right) \rightarrow T_{Reg} + A_{Res} \\ \text{Naive T-cell } \left(T_{Na}\right) + \text{Active Viatmin-D } \left(D^*\right) \rightarrow T_{Reg} + D^* \end{array}\right\} \qquad \text{(vi)}$$

(7) Both regulatory T-cell and active vitamin-D can suppress the production of effector T-cell to control the hyperactivity of the immune system.

$$\left\{\begin{array}{l} T_{Eff} + T_{Reg} \rightarrow T_{Eff}^{\text{killed}} + T_{Reg} \\ T_{Eff} + D^* \rightarrow T_{Eff}^{\text{killed}} + D^* \end{array}\right\} \qquad \text{(vii)}$$

(8) The cycle is completed by the transformation of active APC to resting APC again by the same duo, $T_{Reg}$ and D* which work at tandem.

$$\begin{Bmatrix} A_{Act} + T_{Reg} \rightarrow A_{Res} + T_{Reg} \\ A_{Act} + D^* \rightarrow A_{Res} + D^* \end{Bmatrix} \quad \text{(viii)}$$

We now transform the above set of connected biochemical transformations into rate equations. They give rise to a chemical network.

**B. Master equations quantifying the reaction network dynamics**

Now, we make a few important assumptions before we set about writing the master equations.

(i) Pathogen, inactive APC and naive T-cells, each has a birth rate which includes influx and proliferation rates and a death rate similar to decay which incorporates natural cell death. The death rate of each component is linear in its concentration.

(ii) The transition probabilities are all assumed to be constant with time but may vary from system to system (i.e. here person to person) according to the condition applied.

(iii) To scale the unit, here we assume that in absence of pathogen, hundred (average number of T-cells present in hundred nano-liter blood sample) precursor T-cells pre-exist in our volume element.

The annihilation, recombination and catalytic reactions articulated above lead to the following set of coupled master equations. The equations are size-extensive. Size extensibility is an important element of our model.

$$\frac{dP}{dt} = \sigma_P - m_P P - k_P T_{Eff} P \quad (1)$$

$$\frac{dA_{in}}{dt} = \sigma_A - k_{inp} A_{in} P - m_a A_{in} \quad (2)$$

$$\frac{dA_{Res}}{dt} = k_{inp} A_{in} P + k_{ar} T_{\text{Reg}} A_{Act} + k_{aD*} A_{Act} D* - k_{rese} A_{Res} T_{Eff} - k_{inp} A_{Res} P - m_a A_{Res} \quad (3)$$

$$\frac{dA_{Act}}{dt} = k_{rese} A_{Res} T_{Eff} + k_{inp} A_{Res} P - k_{ar} T_{\text{Reg}} A_{Act} - k_{aD*} A_{Act} D* - m_a A_{Act} \quad (4)$$

$$\frac{dT_{Na}}{dt} = \sigma_T - k_{an} A_{Act} T_{Na} - k_{resn} A_{Res} T_{Na} - k_{nD*} T_{Na} D* - m_n T_{Na} \quad (5)$$

$$\frac{dT_{Eff}}{dt} = k_{an} A_{Act} T_{Na} - k_{er} T_{Eff} T_{\text{Reg}} - k_{eD*} T_{Eff} D* - m_e T_{Eff} \quad (6)$$

$$\frac{dT_{\text{Reg}}}{dt} = k_{resn} A_{Res} T_{Na} + k_{nD*} T_{Na} D* - m_r T_{\text{Reg}} \quad (7)$$

$$\frac{dD_{in}}{dt} = \sigma_D - k_{eD} T_{Eff} D_{in} - m_D D_{in} \quad (8)$$

$$\frac{dD*}{dt} = k_{eD} T_{Eff} D_{in} - m_{D*} D* \quad (9)$$

where the respective terms denote the following quantities :

$k_x \rightarrow$ Transition probability rates (self-evident from equations)

$\sigma_k \rightarrow$ Production rate by body of component $k$,

$m_i \rightarrow$ Overall death rate of component $i$,

$P \rightarrow$ Concentration of Pathogen,

$A_{in}$ → Concentration of inactive antigen presenting cells without pathogen capture,

$A_{Res}$ → Concentration of resting antigen presenting cells after pathogen capture,

$A_{Act}$ → Concentration of activated antigen presenting cells after pathogen recognition and effector T-cell contact.

$T_{Na}$ → Concentration of naive T-cells,

$T_{Eff}$ → Concentration of effector T-cells,

$T_{\mathrm{Reg}}$ → Concentration of regulatory T-cells,

$D_{in}$ → Concentration of Inactive form of Vitamin-D3 (**1, 25(OH)D**) in the body

$D^*$ → Concentration of active form of Vitamin-D3 (**1, 25$(OH)_2$D**) in the body

That is, we have used the same letter to denote both the species and its concentration. This should not cause any confusion.

**C. System parameters and data analysis**

A set of nine coupled differential equations is difficult to solve analytically. We obtain the time dependent concentrations of all the components involved in the scheme by employing the well-known stochastic simulation analysis proposed by Gillespie [62]. Both the single molecular as well as ensemble enzyme catalysis have been studied following this method. All the results presented in this article are derived using stochastic simulation method. However, we have also verified the consistency of each result by using the direct deterministic approach which is easier to implement.

As already mentioned, here we have considered one hundred nano-litre volume of blood sample. In the absence of pathogen this blood sample effectively contains the steady state concentration of all the precursor cells [68-70]. Since all the reactions are bimolecular, the volume dependence of the reaction is expected to be an issue. Thus, we have kept fixed the box volume to one hundred nano-liter and all the rate constants are in the unit of per day.

Furthermore, we have assumed that in the absence of antigen, hundred precursor T-cells can pre-exist within this fixed volume (100 nano-litre), in accord with known experimental values [68, 69]. These T-cells have a 1% turnover per day. Concentrations of pathogens and APCs are also normalized. The production rate and death rate of these components are so assigned that their steady state values become one. Other associated probabilities/rate constants of different reaction sets are used from early papers in this field [44, 45]. However, for vitamin D, the production and mortality rate constants are calculated from their steady state concentration. Other vitamin D related rate constants are treated as variable in our study, as we have no experimental data available for them. For successful implementation of such a model, we require estimates several rate parameters values. Unfortunately accurate values of some of these rate constants are very hard to determine. Such rate parameters depend on several factors and differ from species to species. So they do not have any specific standard value. For example, it would be quite difficult to determine the mortality rate of effector and regulatory T-cell as in the present model these rate parameters also include the proliferation rate along with their death rate. Moreover, the pathogenic stimulation could be of various ranges according to their strength (virulence) and pattern.

Hence the primary difficulty of any predictive theoretical research in this area is the absence of accurate values of rate constants/transition probabilities. In the present study we have employed the following approach to circumvent this difficulty. (i) In some cases where values could be estimated from literature, we have used the known value and varied them over a range to check the sensitivity of results. (ii) In a few cases, order of magnitude estimates for values were employed [44]. We also focused on exploring the phase diagram by varying some key rate parameters that are not known and looked for the optimum region where results are sensitive to the parameter space (given experimental and assumed values of the rate constants and concentrations). To this end, we have varied the rate constants over a significantly wide range. In addition, the concentration of precursor elements was normalized, so as to reflect manifold change in the production level. Taking typical values as mentioned below (see **Table 1**), the time evaluation of the system and other analyses are performed in the present work. Here we have used the standard definition of steady state, i.e.; when the concentration of different species is invariant with time (dc/dt=0). In particular, for stochastic simulation, a steady state is assumed to reach when the concentration of a species fluctuates around a mean value without any noticeable drift at long time.

**Table 1: Basic parameter values (*time duration is taken as "days").**

| Parameter | Symbol | Value |
|---|---|---|
| **Reproduction rate of pathogen** | $\sigma_P$ | **1** |
| **Death rate of pathogen** | $m_P$ | **1** |
| **Birth rate of APC** | $\sigma_A$ | **0.2** |
| **Death rate of APC** | $m_a$ | **0.2** |
| **Rate of pathogen killing by efffector-T cells** | $k_P$ | **100** |
| **Rate of APC activation by pathogen** | $k_{inp}$ | **Variable** |
| **Rate of APC reactivation by effector T cells** | $k_{rese}$ | **Variable** |
| **Rate of APC inhibition by regulatory T cells** | $k_{ar}$ | **$10^{-1}$** |
| **Rate of APC inhibition by active vitamin-D** | $k_{aD*}$ | **$10^{-7}$** |
| **Birth rate of naive T cells** | $\sigma_T$ | **1** |
| **Rate of differentiation of naive T cell to effector T cell induced by active APC** | $k_{an}$ | **1** |
| **Rate of differentiation of naive T cell to regulatory T cell induced by resting APC** | $k_{resn}$ | **1** |
| **Mortality rate of naive T cell** | $m_n$ | **0.01** |
| **Rate of inhibition of effector T cell by active vitamin-D** | $k_{eD*}$ | **0.001** |
| **Rate of inhibition of effector T cell by regulatory T cell** | $k_{er}$ | **10** |
| **Rate of decay of effector T cells** | $m_e$ | **0.1** |
| **Rate of regulatory T cell reactivation by active vitamin-D** | $k_{nD*}$ | **$10^{-7}$** |
| **Rate of decay of regulatory T cells** | $m_r$ | **0.1** |
| **Production rate of inactive vitamin-D** | $\sigma_D$ | **1** |
| **Death rate of inactive vitamin-D** | $m_D$ | **$10^{-9}$** |
| **Rate of reactivation of active vitamin-D induced by effector T cells** | $k_{eD}$ | **$10^{-7}$** |
| **Rate of deactivation of active vitamin-D** | $m_{D*}$ | **$10^{-2}$** |

## IV. Results

Personalized immune system is a *responsive network of dynamical events.* In our mathematical model these events are included through values of rate constants and initial pre-existing concentrations of cells, like APC, naive T-cells, vitamin D concentration. These values characterize the person based diversity in immune system. Here we are interested to explore how different set of parameters control immune response to the invasion by antigens.

### A. Immune response via T-cell fluctuations

A healthy immune system usually functions with a balanced regulation that maintains the population of effector T-cells at an appropriate level, adequate for the clearance of pathogens. The production of effector T-cells again depends on the APC activation process, controlled by the rate parameter of APC activation by pathogenic stimulation ($k_{inp}$). We have studied all the three regulation limits as mentioned before at different pathogenic stimulation ($k_{inp}$). $k_{inp}$ is an important rate parameter in our analysis, and in adaptive immune system.

As discussed earlier, the most stable and safe physiological state is the moderately regulated regime. This state is balanced between the two states, both of which are undesirable. This balance is achieved via a bi-stability which may give rise to large fluctuations that can be harnessed for immune purpose, if and when needed.

There may be experimental evidence on the role of fluctuation in immune response. A recent study conducted with 20 healthy individuals after strenuous physical activity showed significant decrease in the CD4 lymphocyte count after the rest for 60 minutes. The baseline mean CD4+ T cell count in these individuals at 0 min was $1060 x 10^6$/l which was found to decrease to 660x $10^6$/l after a rest for 60 minutes. Hence concentration of CD4+ T cells has been observed to fluctuate significantly in a healthy system [71]. Thus it is natural to assume that such variation is exploited in the immune response.

In order to explore the T-cell population fluctuations in the three regulation regime as mentioned earlier, we monitor the quantity, $\frac{\left\langle (\delta[T(i,j)])^2 \right\rangle}{\left\langle [T(i,j)] \right\rangle}$ varying the rate of pathogenic stimulation, $k_{inp}$. In certain sense, this expression is the analog of the thermodynamic response given by the isothermal compressibility. However, this response is calculated at the steady state condition achieved by a constant, small influx of pathogen.

It is observed that our system of equations display a rather sharp crossover from a strong to weak regulation as the magnitude of pathogenic stimulation ($k_{inp}$) increases from a low value (see **Figure 2**). This crossover is a sensitive function of many parameters, notably the vitamin D concentration. At the crossing point, we observe a bistability in immune response.

Each peak position in the curves forms the ridge shown in **Figure 2**. *Relative fluctuations of T-cells (both for effector and regulatory) exhibit a rapid (divergent-like) growth at a point near the bistable crossover region.* From our results it seems that system requires relevant fluctuation at homeostasis.

A recent clinical study has reported that the disease states in cancer patients both $T_{Eff}$ mean frequency and its mean coefficient of variation are rather significantly reduced [71, 72]. However this study did not find any such change in regulatory T-cell subset. Many other case studies show that cancer patients have a significantly higher frequency of regulatory T-cells. This signifies that the system is under strong regulation. In 2012, a pilot study over a number of ovarian cancer patients revealed the increased levels of recently activated regulatory T-cell subsets with tumor migrating ability (CD4+CD25$^{hi}$Foxp3+CD127−CCR4+CD38+ cells) in those patients when compared to healthy participants [73]. Such enhanced level of regulatory T-cells are largely found to be reluctant to being "re-set" to healthy control homeostatic levels even after the chemotherapeutic perturbation.

In that relation, **Figure 2** shows that not only under strong regulation limit, effector and regulatory T-cell fluctuations decrease at any disease state, whether it is cancer prone or autoimmune. We believe that due to the dearth of necessary fluctuations (especially under the strong regulation limit), regulatory T-cell frequency becomes highly resistant to being "re-set" to their normal homeostatic level in cancer patients, as indeed been observed in some cases.

The present study suggests that not only the absolute values of the effector and regulatory T-cell populations are important, but their fluctuations are also important and independent measure of immune response system.

Our results also clearly demonstrate that for such complex in vivo systems, the occurrence of instabilities and non-equilibrium phase transition phenomena cannot be reliably understood from a complete deterministic approach as such description completely ignores system's inherent fluctuations. The significance of critical fluctuations that abound in the vicinity of a critical point has been repeatedly highlighted in the study of phase transitions. Studies of such collective organizational phenomena often use the concept of scaling laws obeyed by thermodynamic observables close to second-order phase transitions. Strikingly, in the present case such soft fluctuations are too abundant that they may alter the nature of the immune regulation profoundly. More importantly, the system may use these fluctuations in immune response by undergoing phase transition into the other regulation phase. Nevertheless, it is important to note that except some very sensitive order parameter, such isolated fluctuation is rather difficult to arrest in a number of order parameter profiles.

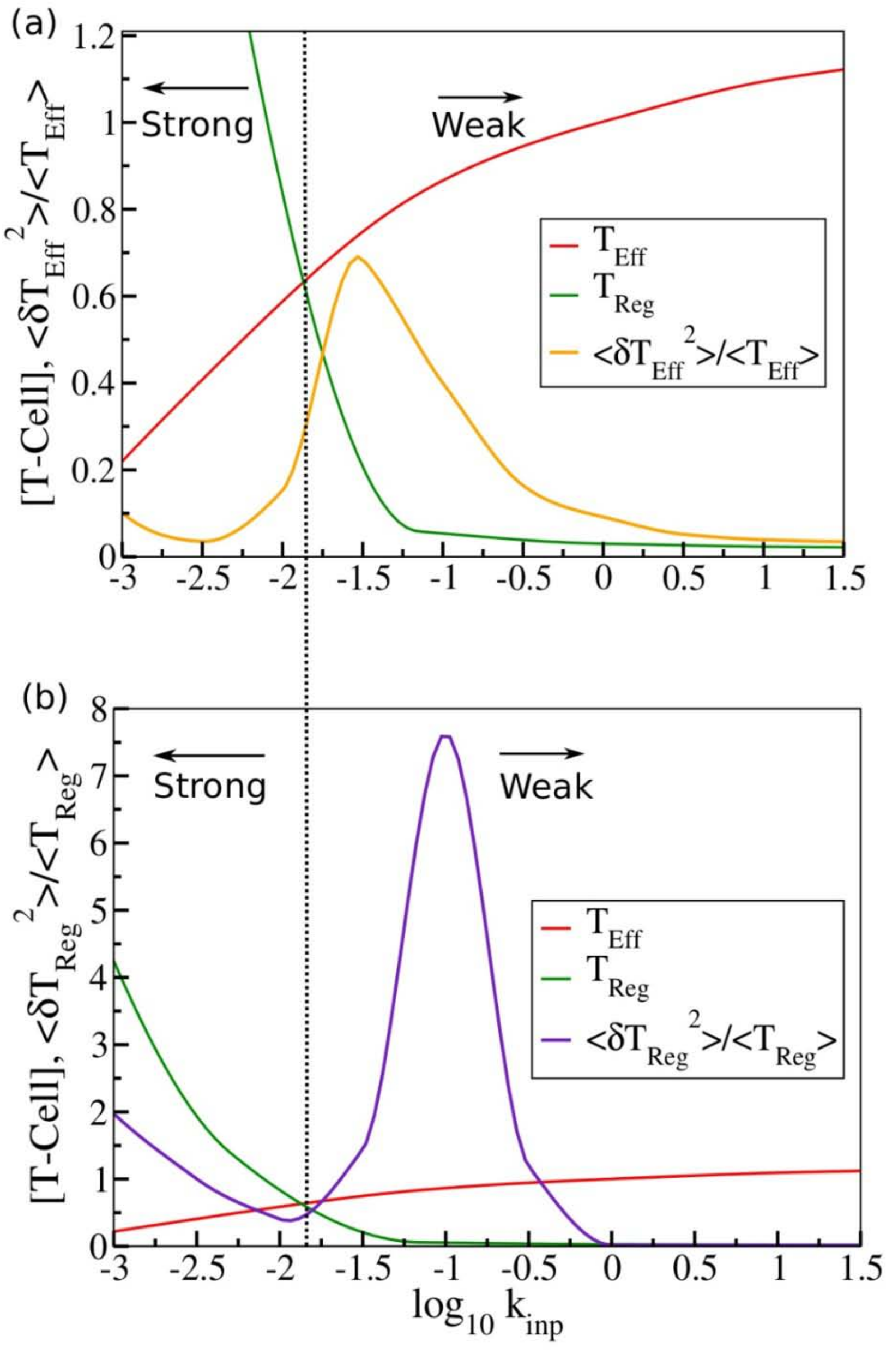

**Figure 2: Relative fluctuation of T-cell population both for (a) effector and (b) regulatory subsets. Both show divergent-like behavior near the crossing of strongly and weakly regulated states. The crossing is marked by a black dashed line where mean population of effector T-cells (red solid line) triumph over the mean population of regulatory T-cells (green solid line). Basic value parameters**

**for this analysis are taken from Table 1. We perform this analysis in presence a standard level of vitamin-D (50nmol/L). Note the change in scale in the two panels.**

## B. Personalized immune response and tolerance

We have discussed earlier that the values of rate constants and pre-existing concentration of inactive cells like APC, naïve T-cells, and inactive vitamin-D together can determine the state of a person's immunity and tolerance. The fact that vitamin D has been implicated as an important factor in several different autoimmune diseases suggests that vitamin D might be an essential element that normally participates in the control of self-tolerance. We show here that tolerance could be intimately connected with bistability of healthy immune system. It is worth mentioning here that experimental observations related to self-tolerance also supports the emergence of bistability where the balanced co-existence of strong and weakly regulated immune response preserve in the system. Note that while vitamin-D assisted strongly regulated state critically boosts up system's tolerance for a certain pathogenic stimulation, weak regulation instigated always acts against that pathogen.

We have investigated variation of regulation that occurs in response to pathogenic stimulation in different immune profiles characterized by different levels of inactive vitamin-D. The results are shown in **Figure 3**. In this analysis we explore how the regulation changes from strong to weakly regulated states causing autoimmunity and escape tolerance. We find that at low concentration (below 50nmol/L) of vitamin-D, system is rather antagonistic towards pathogenic load which leads to a more steady state population of effector T-cells. However, in the presence of higher concentration (beyond 50-100nmol/L) of vitamin D as shown in **Figure 3** we find a considerable enhancement of the tolerance level where system is rather generous to the pathogenic load by suppressing the production of effector T-cell with substantial up-regulation of regulatory T-cell population.

We have monitored normalized mean square fluctuation in the autoimmune prone region and tolerant region. Here we consider a person having less inactive vitamin-D concentration. In that state the response of an immune system is expected to be different from the state where inactive vitamin-D concentration is sufficient. In our model study we indeed find the response defined in terms of fluctuation to diverge near the tolerance limit. *However this divergent-like behavior in the mean square fluctuation for effector T-cell population starts to disappear as we increase the pre-existing concentration level of inactive vitamin-D.* In contrast, such disappearance of divergent-like growth in the same response parameter for regulatory T-cell population is not significant even at higher vitamin-D level.

The present analysis reveals that the observed significant reduction of relative fluctuation of effector T-cell to be a more sensitive diagnostic target to quantify the immune response. The reduction is observed after pathogenic stimulation of a patient living with higher dose of vitamin-D or any other immunosuppressant. Such higher dose of vitamin-D induced immunosuppressive effect imitates the conditions of patients with gynaecological malignancies where the coefficient of variation of effector T-cell was found to be lower in cancer patients compared to healthy donors due to immune suppression [72]. On the contrary, for a patient having relatively lower dose of vitamin-D or any other analogue immunosuppressant, it seems that the relative fluctuation of both effector and regulatory T-cell may serve to be the efficient diagnostic parameters. But again the broader window in the divergence of effector T-cell fluctuation makes the effector cell population fluctuation a more robust response parameter than the regulatory subset in such cases. Moreover, the result emphasizes the scope of diagnosis of a disease in a patient specific manner.

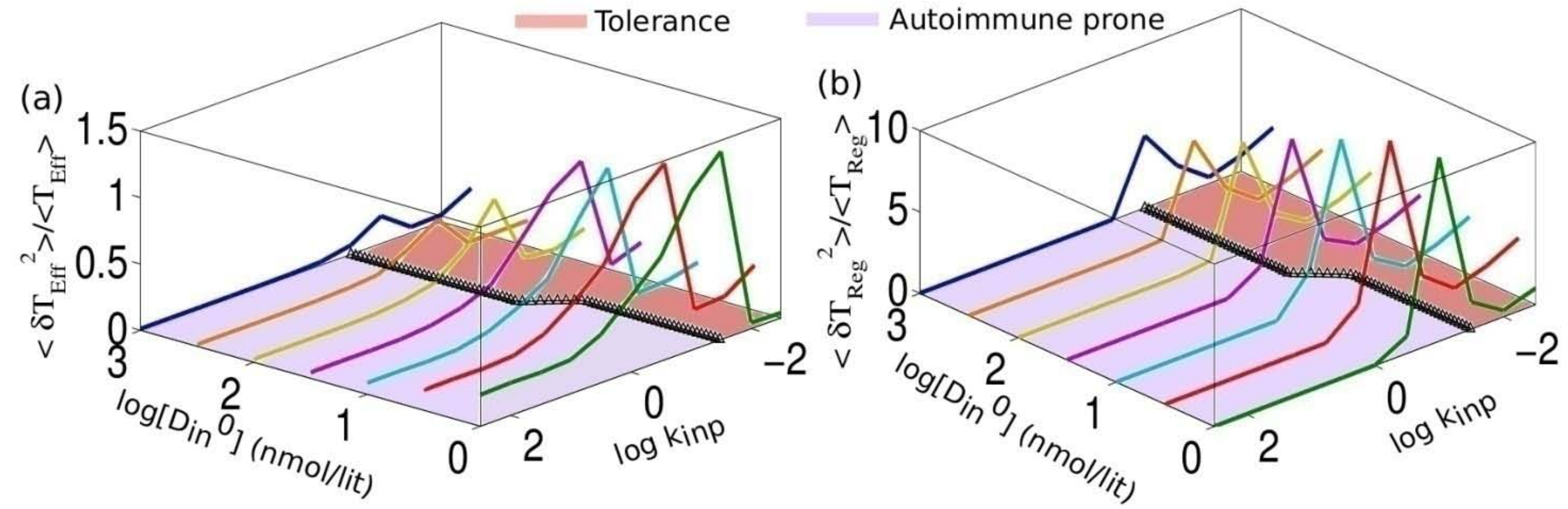

**Figure3: Vitamin-D induced enhancement of tolerance level (emergence of strong regulation). The black line separated orange zone corresponds to the tolerance, i.e. the system falls into the strongly regulated state where it can tolerate certain pathogenic load. Pink zone corresponds to the autoimmune prone region where the system falls into the weakly regulated state. Note that presence of 50-100 nmol/L inactive vitamin-D boosts up system's tolerance. Vitamin-D dependence of immune response in terms of relative fluctuation of regulatory and effector T-cells population are measured along the tolerance and autoimmune prone regions. Relative fluctuation curves show divergent-like behavior near the tolerance limit. At high vitamin-D level (beyond 50-100 nmol/L) relative fluctuation of effector T-cell diminishes while the same for regulatory T-cell does not. At low vitamin-D level (below 50-100 nmol/L) relative fluctuation of effector T-cell has a much broader window of relative fluctuation than the same for regulatory T-cell. Basic value parameters for this analysis are taken from Table 1.**

## C. Bistablity and co-existence of steady state branches

In the light of our previous results, it is worth mentioning here that fluctuations are always present in cellular systems due to the stochastic nature of the elementary processes[48]. These fluctuations exert more profound effect and can induce new activities in small scale systems which may not be predicted or captured by any mean-field description or macroscopic rate laws. While noise, whether external or internal can cause a finite system to fluctuate between two stable states in bistable systems, macroscopic rate laws predict an asymptotic behavior that ignores such oscillations. Hence we need to understand the concept of bistability in the fluctuating environment as it is a basic phenomena of cellular functioning. In our model bistability is augmented in the system due the inhibitory effect of regulatory T cells on the activation of APC and effector T-cells. Furthermore, this augmentation is dependent on the level of vitamin-D. Interestingly, in the bistability condition system becomes more robust to tolerate significant changes of pathogenic stimulation. Here one typically finds strongly broadened probability distribution with a clear bimodal nature that is seen only in the vicinity of bistable point.

As mentioned, we find the occurrence of bistability in the moderate (that is, healthy) regulation regime through the coexistence of two stable steady states as shown in **Figure 4**. When such a system enters the bistable state with roughly equal levels of key determinants (in our case effector and regulatory T-cell populations), effects of stochastic fluctuation become amplified. It also depends on the external rate parameters which may enforce a final selection. As a result of these fluctuations, such complex biological systems may undergo modifications of their steady state properties. In particular, when the variance of fluctuations increases around a well defined mean value, transient phenomena appear which may absent in a typical bifurcation diagrams.

Hence the properties of the non-fluctuating systems may not be considered as a first approximation to the properties of the real system [48].

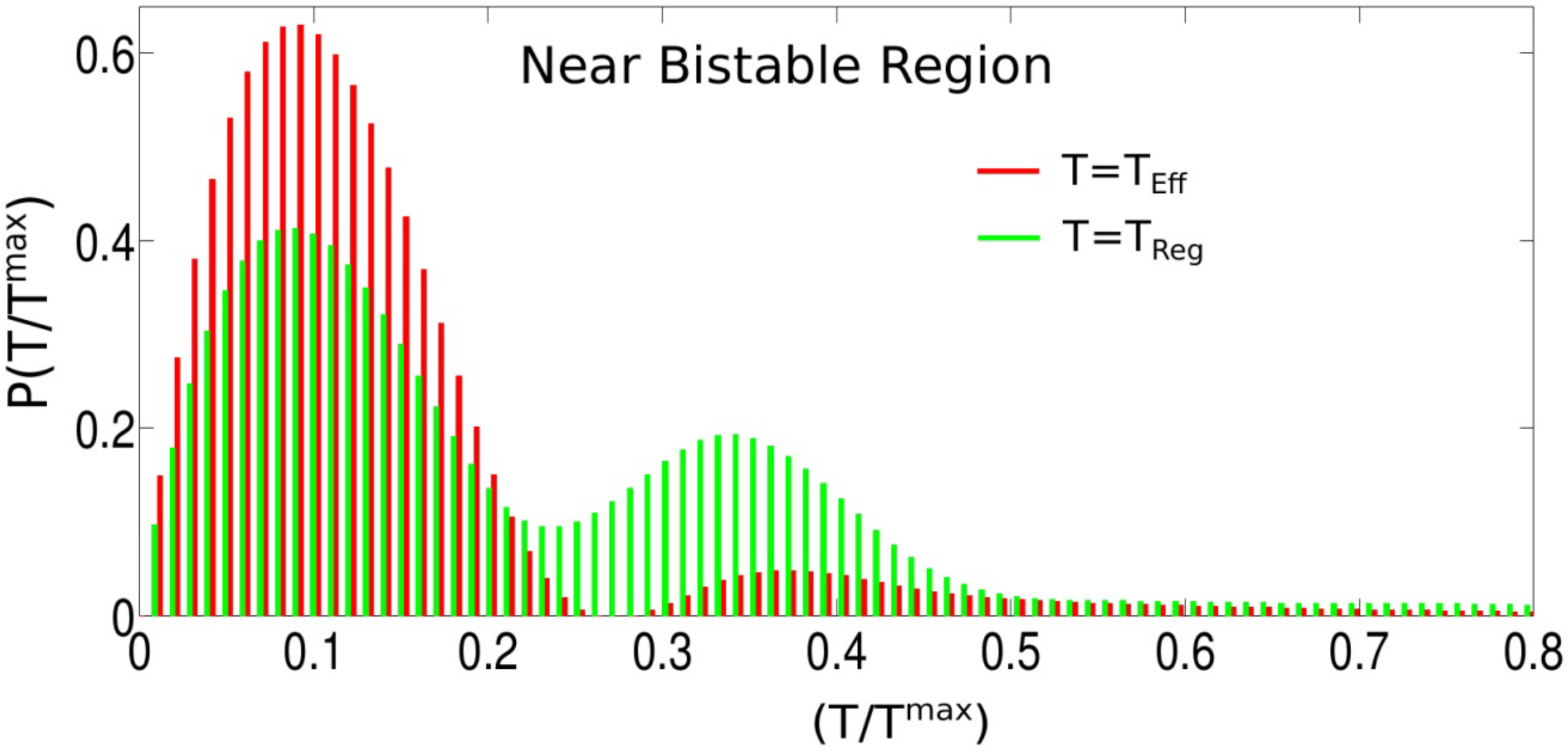

**Figure 4: Population distribution of effector ($T_{Eff}$) and regulatory ($T_{Reg}$) T-cells near bistable region. Both the T-cells show a clear bimodal distribution and are seen only in the vicinity of bistable region. The bimodal nature of the distribution corresponds to the existence of two stable states: weakly regulated state and strongly regulated state. We have normalized both the values of T-cell population by their respective max values for better comparison. $k_{inp}$, $k_{rese}$ are so chosen that they remain in the bistable region ($k_{inp}$=0.1, $k_{rese}$ =100; vitamin-D level ~ 50nmol/L). Other parameters remain unchanged as in Table 1.**

## D. Time-dependent oscillatory behavior o f antigen-specific effector ($T_{Eff}$) and regulatory ($T_{Reg}$) T cells and immune-suppressor dependence

Whether the pathogenic stimulation strong or weak a healthy immune network always maintains a balanced regulation over the effector T-cell concentration to reduce the risk of autoimmunity or cancer. However there are increasing evidences where to treat autoimmune patients, to transplant organ the applications of immunosuppressive drugs are emerging to a greater extent, sometimes without caring their plausible adverse effects. Such immunosuppressive drugs enhance the risks of malignancy, infection, cardiovascular disease and bone marrow suppression. Most importantly immunosuppressive and chemotherapeutic drugs at their high dose have a rather long term (in terms of years) effects and the adverse effects of these drugs may arise even long after the treatment has stopped [74-76].

Here we find that the noise arising from T-cell dynamics could play an important role in controlling the bistablity in presence of pathogen. In particular, the cross-regulation of effector T-cells and regulatory T-cells generates a stable oscillatory dynamics that maintains body's homeostasis. This dynamical cross-regulation has a clinical relevance in promoting the relapsing-remitting events under autoimmune conditions. We observe that the coupled oscillation of effector T-cells ($T_{Eff}$) and regulatory T-cell ($T_{Reg}$) initiates within 2-5 days and periodically continues (see **Figure 3**). It is important to note that their variations move in an anti-correlated fashion.

In an interesting recent article Martinez-Pasamar et al. noted that the generation of an autoimmune response appears to be influences by the presence of susceptibility factors such as genetic polymorphisms, levels of modulators of the immune suppressors such as vitamin D, or previous infections that may change the thresholds of T-cell activation and differentiation, defining the autoimmune regime [77]. To arrest the effects of immune modulators like vitamin-D we have introduced their important effects in our kinetic model. In our early study we have indeed showed for the 1st time that how the level of immune modulators like vitamin-D or any other immune-suppressants can modify the immune regulation regime. In the present study, we use the same kinetic framework in an aim to address specific immune responses and the modulator dependence.

In **Figure 5** we show a long time series demonstrating the dynamical pattern in the population of the two T-cells at three different concentrations of vitamin-D to monitor its dose dependent immunosuppressive effects. The values of other parameters are kept fixed. We observe that in the presence of optimal level of vitamin-D (50nmol/L), system remains in the bistable state where the frequent visits to both the two regulation limits (strong and weak) are clearly reflected in the population dynamics of effector and regulatory T-cells (see **Figure 5**). In contrast, the *system scarcely visits the weak regulation state when we increase the pre-existing level of vitamin-D to its higher dose (200nmol/L)*. However, when we set the pre-existing level of vitamin-D to a significantly lower value (10nmol/L), the system retains the weakly regulated state in most of the occurrences. In fact, when the system tends towards a slightly weak/strong regulation regime from the bistable condition, we observe a less frequent dynamic cross regulation in the temporal progression of regulatory and effector T-cells populations. This could provide a dramatic signature of disease phenotype in clinical therapy [77, 78].

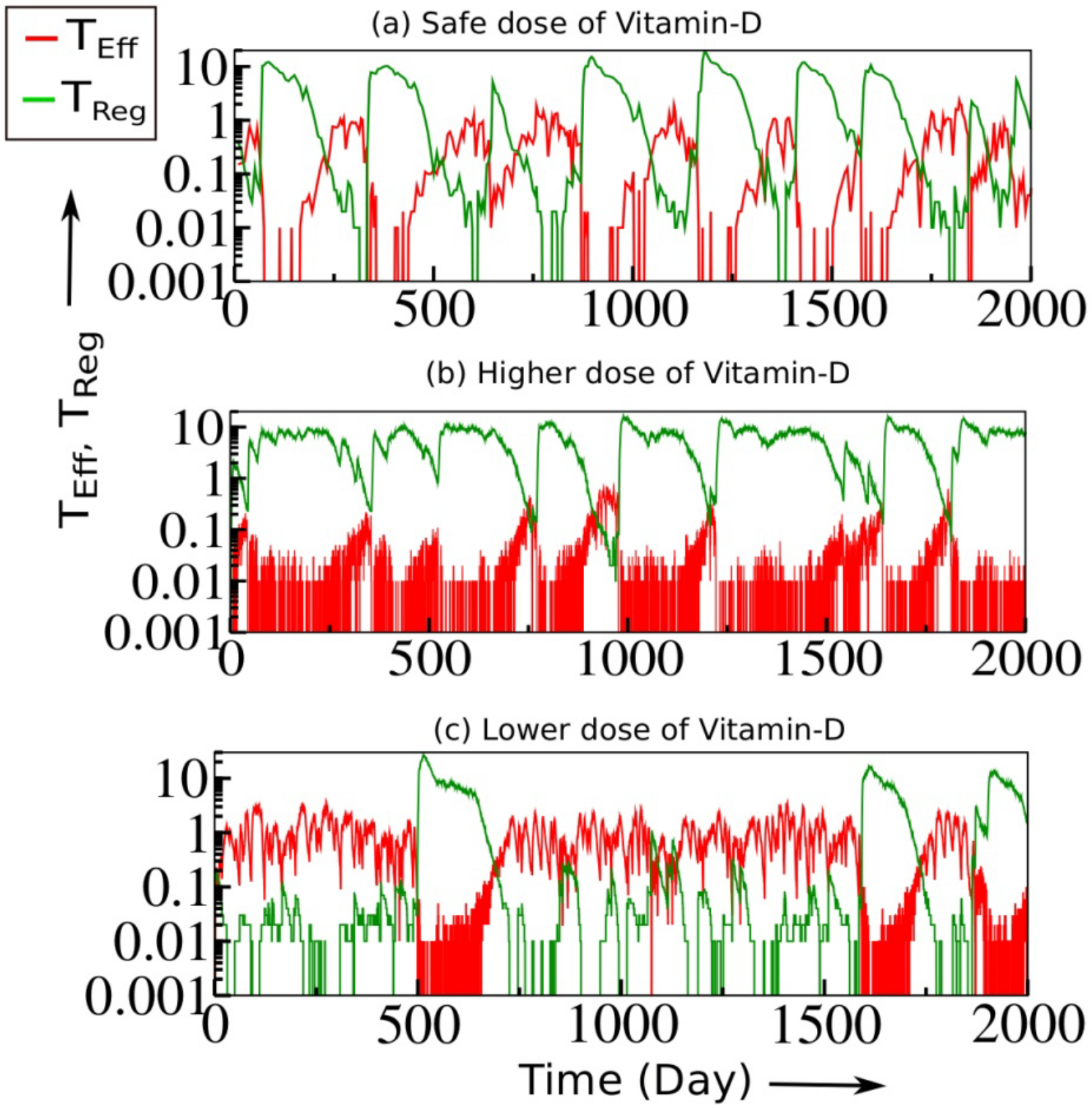

**Figure 5: Impact of vitamin-D over the dynamical profile of effector and regulatory T-cell in presence of pathogen. (a) In presence of optimal level of vitamin-D (50 nmol/L) system visits both the regulation states (weak and strong) quite frequently. Effector T-cells and regulatory T-cells propagate with an oscillation. Note the strong anticorrelated fluctuations between effector and regulatory T-cell subset and intermittency in the fluctuations of both species which are relevant for maintaining the bistability. (b) In presence of higher than optimal level of vitamin-D (200 nmol/L) system almost always resides in the strongly regulation state. (c) In presence of lower than optimal level of vitamin-D (10 nmol/L) system strongly avoids to be in the strongly regulation state. The pattern of such dynamical profiles has experimental and clinical relevance. The rate parameters considered here, are similar to Figure 4 given in main text.**

### E. Relaxation behavior in healthy and disease states

By using the universality observed in critical phenomena it may be possible to provide quantitative predictions for the immunological condition of a given person by following the relaxation behavior [78, 79]. In the present case one can however perform a perturbation (pulse) experiment in situations where the immune system is (i) far from a transition (strong regulation and weak regulation) and (ii) in the case when the state is close to a transition (near bistable region). Such an "experiment" could clearly reveal the difference in time evolution leading to relaxation to its initial equilibrium state. The longer that time the system would take, the less stable is the system and thus we can assure that it is more close to a potential tipping point.

In **Figure 6 (a)** and **(c)** we observe the relaxation behavior of both effector and regulatory T-cell fluctuations respectively. In the present case we indeed find that the relevant properties of an immune system can be determined by the fluctuations of the order parameter. At bistable region their relaxation pattern shows critical phenomena like behavior. Relaxations in weak and strong regulatory regions also have their own characteristic features reflecting relatively faster decay.

We recall **Figure 5** where the temporal variation of the effector and regulatory T-cell near bistable region clearly exhibits intermittency. The signature of intermittent behavior can be confirmed by calculating the power spectrum of the fluctuation time correlation function (TCF) (TCF are shown in **Figure 6(a)** and **(c)**). Power spectrums of the fluctuation of the effector and regulatory T-cell are presented in **Figure 6 (b)** and **(d)** for the three regulation regime, i.e., strong, moderate and weak. The linear dependence of the power spectrum for a wide range of frequency starting from low frequency region with slope $\sim -1$ (in log-log plot) in the precise regime of moderate and weak confirmed the strong intermittency present in the fluctuation for both effector and regulatory T-cell [78]. However at the strongly regulated region the power spectrum only for effector T-cell shows a linear dependence at a rather higher frequency region, indicating the presence of small time scale intermittency in the fluctuations. Thus absence of intermittency in the fluctuation of the population of T-cells in the system may cause a weakening of immune response. Such anomalous response, in turn, may reflect symptoms of a particular disease.

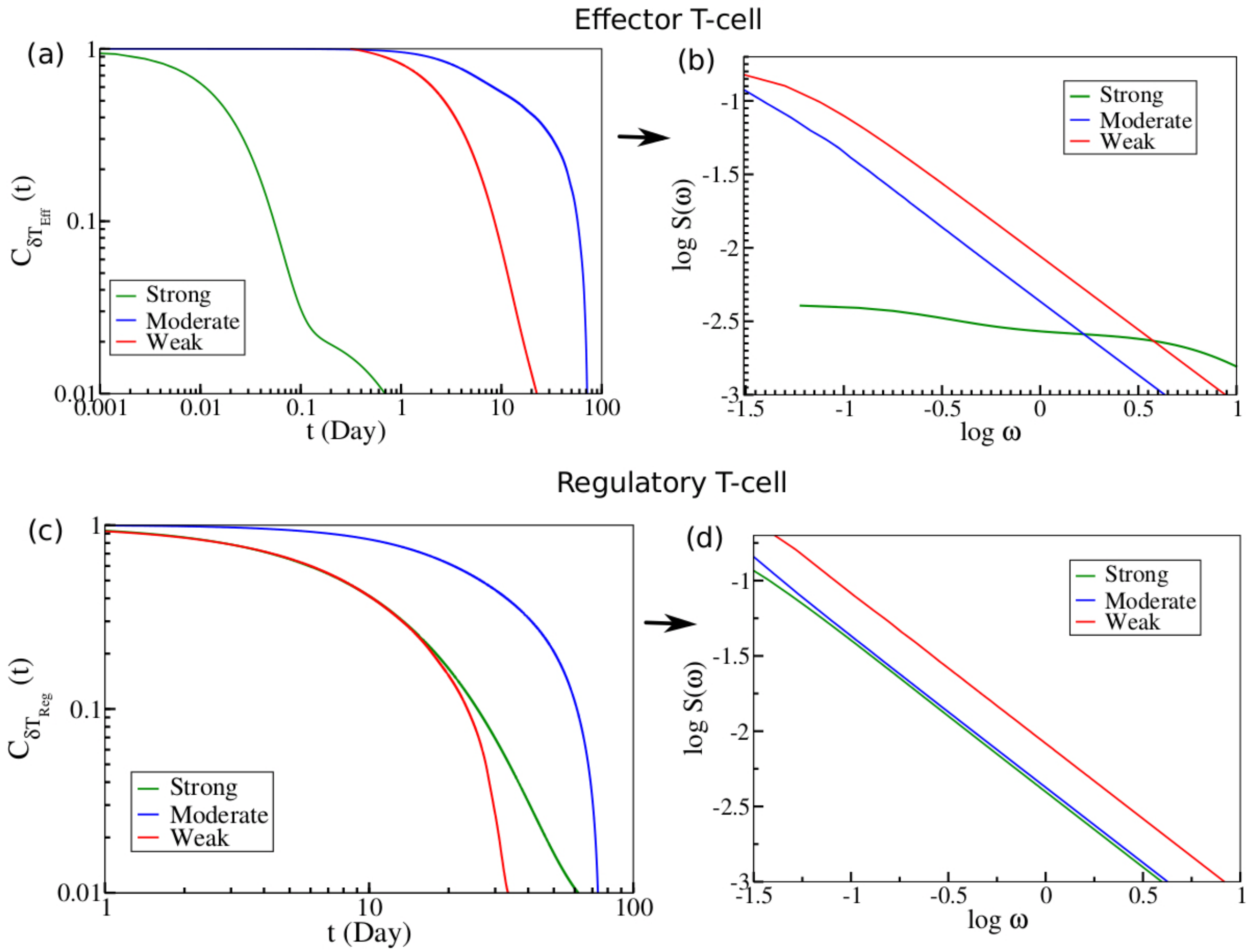

**Figure 6: Relaxation behavior and intermittency in the population fluctuation of effector and regulatory T-cell subset in strong, moderate and weak regulation zones respectively. (a) Time correlation function (TCF) of the fluctuation of effector T-cell population and (b) the power spectrum of the corresponding time correlation function for effector T-cell in log-log plot. (c) TCF of the fluctuation of regulatory T-cell population and (d) the power spectrum of the corresponding time correlation function for regulatory T-cell. For both types of T-cells, the associated TCFs show characteristic slow relaxation behavior at moderate regulation region. At strongly regulated region effector T-cell dynamics is considerable faster. Note the linear dependence over a wide range of frequency with slope $\sim -1$ (distinctive of 1/f noise) for both effector and regulatory T-cells at bistable and weakly regulated conditions, which confirms the large time scale strong intermittency in the fluctuations. The power spectrum for effector T-cell at strongly regulated region shows a straight line with slope $\sim -1$ that appears at much higher frequency ($\omega$) range, which indicates the presence of intermittency in the short time scale.**

## V. Conclusion

In the present work we have attempted to provide a quantitative definition of personalized immune response of an individual human being to a given specific pathogen. We propose that personalized immune response can be quantified by an appropriate response function. Ideally, one should be able to categorize and quantify specific immune response so that appropriate preventive measures can be taken. The application of linear response theory, in this aspect, might be useful in the diagnosis of different stages of a disease in addition to the measurements of ratios or standardization of the concentrations of some potent cells or protein levels currently being advocated [72, 80, 81].

We have followed methods of statistical mechanics in defining such individual-specific immune response functions. *We propose that such response functions can be defined in terms of fluctuations of the relevant concentration of some key determinants*. Progressive scientific researches in the present field have evoked the idea of stochastic bifurcation with noise-induced bistability and its relevance towards nonlinear stochastic biochemical systems [80-82]. Recently Qian et al. [85-87] analyzed a number of biochemical network models and showed that noise-induced bistability in general arises from slow fluctuations, and a pitchfork bifurcation occurs as the rate of fluctuations decreases. Lefever et al. reported the importance of fluctuation modulating steady state properties of biochemical processes drastically and the related implications in tumor immunology [48, 49, 88].

In the present study we have employed a chemical albeit simple network model of immune system to calculate the response from fluctuations of the effector T-cell concentration after pathogenic incursion. As this is coupled to regulatory T-cell concentration, we also calculate fluctuation in the regulatory T-cell concentration. *The important outcome is that while absolute values of both the T-cell concentrations appear within the balanced range in an apparently healthy human being, fluctuations in the concentration of such cells show a divergent like behavior* [72]. Interestingly, both the two quantities show signatures at the boundary between strong and weak regulatory regimes. Fluctuation-driven transitions between strong and weakly regulated branches are found to occur sufficiently close to the bifurcation point. Thus sufficient reduction in fluctuation of a key determinant like effector or regulatory T-cell can be used to be an effective marker for defining a disease state.

The fact that vitamin D has been implicated as an important factor in several different autoimmune diseases suggests that vitamin D might be an essential element that normally participates in the control of self-tolerance. However, we find that in the presence of vitamin D a considerable enhancement of the tolerance level which would lead to a reduced risk of autoimmunity. We also monitor the effects of other susceptibility factors like levels of modulators of the immune response, such as vitamin D on the thresholds of T-cell population in the perspective of various autoimmune diseases. For instance, we can consider a person having less inactive vitamin-D concentration. In that state the response of an immune system is expected to be different from the state where inactive vitamin-D concentration is sufficient. As a case study, we vary pre-existing inactive vitamin-D concentration in a wide scale from its standard homeostatic level. As a consequence we found a drastic reduction in effector T-cell fluctuation upon increased pre-existing vitamin-D level. *The study also demonstrated that while the presence of optimal level of vitamin-D assists to broaden the tolerance range than its lower concentration limit, the range and intensity of essential fluctuation of effector T-cell concentration substantially falls down in the overdose limit.* The bimodal distribution, in presence of optimal concentration of vitamin-D, provides evidence of the fluctuation induced transition between strongly and weakly regulated states in bistable condition. To investigate the dynamical pattern near the bistable condition and also at a condition created by slightly shifting towards the extremes by varying doses of vitamin-D, we observe less frequent visits of one of the stable states (than the other) during the infectious period at the disease states.

At the bistable condition, the strong anticorrelated fluctuations between effector and regulatory T-cell subsets and the associated intermittency in the fluctuations can be effectively used to probe both the disease phenotype and the response to immunotherapy. The power spectrum of the fluctuation time correlation function of T-cell dynamics reflects the intermittent behavior in their dynamics. The substantial faster decay in the relaxation mode of effector T-cell fluctuation at strongly regulated condition provides a marked signature of adaptive immune efficacy while we do not find such rapid drop in the relaxation of regulatory T-cell fluctuation. Such relaxation behaviors show that fluctuation induced bistability arises from slow fluctuations, and bistabilty terminates as the rate of fluctuations decreases. The broad fluctuation profile and significant reduction in the relative fluctuation of effector T-cell in the presence of the immune-modulator like vitamin-D suggest that the fluctuation of effector T-cell to be a more responsive one.

Our model attempts to quantify a response parameter in terms of T-cell fluctuation. While steady state fluctuation of T-cells might help to evoke personalized immunization their characteristic dynamics indicates that some underlying processes are greatly randomized when birth or death occurs in large bursts. Each birth or death of immune component could involve several small steps, creating a memory between the individual events. We hope to identify some of those complex channels by addressing the temporal cooperatively involved in memory preservation in recent future.

## Acknowledgment

SR thanks Milan Kumar Hazra, Tuhin Samanta, Rajesh Dutta, Puja Banerjee and Dr. Biman Jana for many helpful discussions. This work was supported in parts by grants from DST, India. BB acknowledges support from JC Bose fellowship from DST, India.

## References

1. D. A. Beard and H. Qian, *Chemical Biophysics: Quantitative Analysis of Cellular Systems* (Cambridge University Press, 2008).
2. T. J. Kindt, R. A. Goldsby, B. A. Osborne, and J. Kuby, *Kuby Immunology* (Edition: 6, illustrated, Publisher: W. H. Freeman, 2007).
3. C. Waltenbaugh, T. Doan, R. Melvold, and S. Viselli, *Immunology. Lippincott's Illustrated reviews* (Philadelphia: Wolters Kluwer Health/Lippincott Williams & Wilkins, 2008).

4. E. Kirkwood and C. Lewis, *Understanding Medical Immunology* (John Wiley & Sons, Chichester, 1989).
5. G. Virella, *Medical Immunology*. (6th Edition, New York, Basel: Marcel Dekker, 2001).
6. G. MacPherson and J. Austyn, *Exploring Immunology: Concepts and Evidence* (Wiley-Blackwell, 2012).
7. P. J. Delves and I. M. Roitt, *N. Engl. J. Med.* **343**, 37-49 (2000).
8. B. Alberts, J. Alexander, L. Julian, R. Martin, R. Keith, and W. Peter, *Molecular Biology of the Cell* (Fourth Edition. New York and London: Garland Science, 2002).
9. G. Mayer, *Immunology Section of Microbiology and Immunology On-line*, (University of South Carolina).
10. C. A. Janeway, Jr, P. Travers, M. Walport, and M. J. Shlomchik *Immunobiology, The Immune System in Health and Disease*, (5th edition, New York: Garland Science; 2001).
11. M. T. Lotze and K. J. Tracey, *Nature reviews. Immunology* **5**, 331-42 (2005).
12. C. A. Janeway, Jr. *Immunobiology* (6th ed., Garland Science, 2005).
13. S. Arai, R. Meagher, M. Swearingen, H. Myint, E. Rich, J. Martinson, H. Klingemann, C*ytotherapy* **10**, 625-32 (2008).
14. L. Adorini, *Cellular Immunology* **233**, 115-124 (2005).
15. S. Gregori M. Casorati, S. Amuchastegui, S. Smiroldo, A. M. Davalli, and L. Adorini, *J Immunol.* **167**, 1945-1953 (2001).
16. S. Sakaguchi, N. Sakaguchi, M. Asano, M. Itoh, and M. Toda, *J. Immunol*. **155**, 1151-1164 (1995).

17. C. W Chan, E. Crafton, H.-N. Fan, J. Flook, K. Yoshimura, M. Skarica, D. Brockstedt, T. W. Dubensky, M. F. Stins, L. L. Lanier, D. M. Pardoll, and F. Housseau. *Nat Med.* **12**, 207-13 (2006).

18. J. R. Cochran, T. O. Cameron, J. D. Stone, J. B. Lubetsky, and L. J. Stern *J. Biol. Chem.* **276**, 28068-28074 (2001).

19. J. Huang, V. I. Zarnitsyna, B. Liu, L. J. Edwards, N. Jiang, B. D. Evavold, and C. Zhu, *Nature* **464** 932 (2010).

20. A. Jäger and V. K. Kuchroo, *Scand J Immunol.* **72**, 173-184 (2010).

21. L. J. Edwards, V. I. Zarnitsyna, J. D. Hood, B. D Evavold, and C. Zhu, *Front. Immunol.* **3**, 86 (2012).

22. I. Gutcher and B. Becher, *J. Clin. Invest.* **117**, 1119-1127 (2007).

23. T. Willinger, T. Freeman, H. Hasegawa, A. J. McMichael, and M. F. Callan, *Journal of Immunology* **175**, 5895-903 (2005).

24. M. T. Cantorna and B. D. Mahon, *Exp. Biol. Med. (Maywood)* **229**, 1136-42 (2004).

25. T. R. Mosmann and R. L. Coffman, *Annu. Rev. Immunol.* **7**, 145-73 (1989).

26. H. Patt, T. Bandgar, A. Lila, and N. Shah, *Indian. J. Endocrinol. Metab.* **17**, s612-s61 (2013).

27. T. Rhen and J. A. Cidlowski, *N. Engl. J. Med.* **353**, 1711-1723 (2005).

28. D. C. Hsu and C, H. Katelaris, *Aust Prescr* **32**, 68-71(2009).

29. M. A. Moyad, *Dermatology Nursing* **21**, 1-11 (2009).

30. R. Z. Stolzenberg-Solomon, E. J. Jacobs, A. A. Arslan, D. Qi, A. V. Patel, K. J. Helzlsouer, S. J. Weinstein, M. L. McCullough, M. P. Purdue, X.-O. Shu, K. Snyder, J. Virtamo, L. R. Wilkins, K. Yu, A. Zeleniuch-Jacquotte, W. Zheng, D. Albanes, Q. Cai, C. Harvey, R. Hayes, S. Clipp, R. L. Horst, L. Irish, K. Koenig, L. L. Marchandand, and L. N. Kolonel, *Am. J. Epidemiol.* **172**, 81-93 (2010).

31. D. L. Kamen and V. Tangpricha, *J Mol Med (Berl)* **88**, 441–450 (2010).

32. D. Bikle, *J. Clin. Endocrinol. Metab.* **94**, 26-34 (2009).

33. S. Dimeloe, A. Nanzer, K. Ryanna, and C. Hawrylowicz, *J. Ster. Biochem. Mol. Bio.* **120**, 86-95.

34. G. Penna, A. Roncari, S. Amuchastegui, K. C. Daniel, E. Berti, M. Colonna, and L. Adorini, *Blood* **106**, 3490-3497 (2005).

35. S. Gregori, M. Casorati, S. Amuchastegui, S. Smiroldo, A. M. Davalli, and L. Adorini, *J. Immunol.* **167**, 1945-1953 (2001).

36. I. Schuster, H. Egger, G. Herzig, G. S. Reddy, J. A. Schmid, M. Schüssler, and G. Vorisek, *Anticancer Res.* **26**, 2653-2668 (2006).

37. J. Correale, M. C. Ysrraelit, and M. I. Gaitán, *Brain* **132**, 1146-1160 (2009).

38. E. Peelen, S. Knippenberg, A. H. Muris, M. Thewissen, J. Smolders, J. W. C. Tervaert, R. Hupperts, and J. Damoiseaux, *Autoimmunity Reviews* **10**, 733-743 (2011).

39. H. Jonuleit, E. Schmitt, K. Steinbrink, A. H. Enk, *Trends Immunol.* **22**, 394-400 (2001).

40. M. B. Lutz and G. Schuler, *Trends Immunol.* **23**, 445-449 (2002).

41. H. Jonuleit, E. Schmitt, G. Schuler, J. Knop, and A. H. Enk, *J. Exp. Med.* **192**, 1213-1222 (2000).

42. F. Powrie and K. J. Maloy, *Science* **299**, 1030-1031 (2003).

43. L. Piemonti, P. Monti, M. Sironi, P. Fraticelli, B. E. Leone, E. D. Cin, P. Allavena, and V. D. Carlo, *J. Immunol.* **164**, 4443-4451.

44. S. Roy, K. Shrinivas, and B. Bagchi, *PLoS ONE* **9**, e100635 (2014).

45. D. Fouchet and R. A. Regoes, *PLoS ONE* **3**, e2306 (2008).

46. A. Park, C. Govindaraj, S. D. Xiang, J. Halo, M. Quinn, K. Scalzo-Inguanti, M. Plebanski, *Cancers* **4**, 581-600 (2012).

47. C. C. Preston, M. J. Maurer, A. L. Oberg, D. W. Visscher, K. R. Kalli, L. C. Hartmann, E. L. Goode, K. L. Knutson, *PLoS ONE* **8**, e80063 (2013).

48. R. Lefever and W. Horsthemke, *Bull Math Biol.* **41**, 469-90 (1979).

49. R. Lefever and T. Erneaux, *Nonlinear Electrodynamics in Biological Systems* (Spring link, ISBN: 978-1-4612-9720-8, 1984, 287-305).

50. R. Kubo, *Journal of the Physical Society of Japan*, **12**, 570-586 (1957).

51. J. P. Sethna, *Chapter 10: Correlations, response, and dissipation". Statistical Mechanics: Entropy, Order Parameters, and Complexity* (Oxford University Press. ISBN 0198566778, 2006).

52. R. Kubo, M. Toda, and N. Hashitsume, *Springer Series in Solid-State Sciences* **31**, 146-202 (1991).

53. B. Bagchi, *Molecular Relaxation in Liquids* (Oxford University Press, USA, 2012).

54. H. Hansen and I. R. McDonald, *Theory of Simple Liquids* (Academic, London, 1986).

55. J. J. Hopfield, *Proc. Nat. Acad. Sci. USA* **71,** 4135 (1974).

56. H. Qian, Biophys. Chem. 105,585-93 (2003).

57. S. Chaudhury , J. Cao, and N. A. Sinitsyn, *J. Phys. Chem. B* **117**, 503-509 (2013).

58. Z. Wu and J. Xing, *Biophys. J.* **103**, 1052-1059 (2012).

59. W. Min, X. S. Xie, and B. Bagchi, *J. Phys. Chem.* **112,** 454-466 (2008).

60. M. Santra and B. Bagchi, *J. Phys. Chem. B* **116,** 11809-11817 (2012).

61. M. Santra and B. Bagchi, Plos One 8, e66112 (2013).

62. D. T. Gillespie, *J. Comput. Phys.* **22**, 403–434 (1976).

63. T. E. Turner, S. Schnell, and K. Burrage, *Comput. Biol. Chem.* **28**, 165-78 (2004).

64. V. Mandlik, S. Shinde, A. Chaudhary, and S. Singh, *Integr. Biol.* **4**, 1130-1142 (2012).

65. N. K. Jerne, *Ann Immunol (Paris)* **125C**, 373-89 (1974).

66. J. W. Pike, M. B. Meyer, and K. A. Bishop, *Rev. Endocr. Metab. Disord.* **13**, 45–55 (2012).

67. J. C. Fleet, *J. Nutr*. 134, 3215–3218 (2004); J. C. Fleet, *Mol. Aspects Med*. **29**, 388–396 (2008).

68. G. J. Ligthart, H. R. Schuit, and W. Hijmans, *Immunology* **55**, 15-21 (1985).

69. M. F. Bates, A. Khander, S. A. Steigman, T. F. Tracy Jr., and F. I. Luks, *Pediatrics* **133**, e39-44 (2014).

70. D. B. Fearnley, L. F. Whyte, S. A. Carnoutsos, A. H. Cook, and D. N. Hart, *Blood* **93**, 728-36 (1999).

71. A. Shete, M. Thakar, P. R. Abraham, and R. Paranjape, *Indian J. Med. Res.* **132**, 667-675 (2010).

72. M. T. Madondo, S. Tuyaerts, and B. B.Turnbull, *J. Transl. Med.* **12**, 179 (2014).

73. A. Park, C. Govindaraj, S. D. Xiang, J. Halo, M. Quinn, K. Scalzo-Inguanti, and M. Plebanski, *Cancers* **4**, 581-600 (2012).

74. D. C. Hsu and C. H. Katelaris, *Aust Prescr.* **32**, 68-7 (2009).

75. L. M. Shaw, B. Kaplan, and K. L. Brayman, *Clin. Chem.* **44**, 381-7 (1998).

76. F. F. Fagnoni, L. Lozza, C. Zibera, A. Zambelli, L. Ponchio, N. Gibelli, B. Oliviero, L. Pavesi, R. Gennari, R. Vescovini, P. Sansoni, G. Da Prada, G. Robustelli Della Cuna, *Immunology* **106**, 27-37 (2002).

77. S. Martinez-Pasamar, E. Abad, B. Moreno, N. V. D. Mendizabal, I. Martinez-Forero, J. Garcia-Ojalvo, and P. Villoslada, *BMC Systems Biology* **7**, 34 (2013).

78. N. Figueroa-Moralesa, K. Leónc, and R. Muleta, *J. Theo. Biol.* **295**, 37–46 (2012).

79. L. Mollet, T-S. Li, A. Samri, C. Tournay, R. Tubiana, V. Calvez, P. Debré, C. Katlama, B. Autran, and the RESTIM and COMET Study Groups, *J. Immunol.* **165**, 1692-1704 (2000).

80. R. Chowdhury, B. Jana, A. Saha, S. Ghosh, and K. Bhattacharyya, *Med. Chem. Commun.* **5**, 536-539 (2014).

81. S. Chattoraj, S. Saha, and S. S. Jana, and K. Bhattacharyya, *J. Phys. Chem. Lett.*, **5**, 1012-1016 (2014).

82. R. Zwanzig, *J. Chem. Phys.* **97**, 3587 (1992).

83. R. Zwanzig, *Acc. Chem. Res.* **23**, 148-152 (1990).

84. J. E. M. Hornos, D. Schultz, G. C. Innocentini, J. Wang, A. M. Walczak, J. N. Onuchic and P. G. Wolynes, *Phys. Rev. E* **72**, 051907–5 (2005).

85. H. Qian, P.-Z. Shia, and J. Xing, *Phys. Chem. Chem. Phys.* **11**, 4861-4870 (2009).

86. H. Qian, *Biophys. Chem.* **105**, 585-593 (2003).

87. H. Ge and M. Qian, *J. Chem. Phys.* **129**, 015104 (2008).

88. W. Horsthemke and R. Lefever, *Noise-Induced Transitions, Theory, Apllications in Physics, Chemistry, and Biology* (Springer Series in Synergetics, Springer-Verlag, Berlin, 1984, vol. 15).